\documentclass{openjournal} 

\usepackage{graphicx}
\usepackage{rotating}
\usepackage{txfonts}
\newcommand{\degree}{\ensuremath{^\circ}}

\begin{document}

   \title{A targeted, parallax-based search for Planet Nine}


 \author{Hector Socas-Navarro}
        \affiliation{European Solar Telescope Foundation, V\'\i a L\'actea S/N, La Laguna 38205, Tenerife, Spain}
 
        \affiliation{Instituto de Astrof\'\i sica de Canarias, V\'\i a L\'actea S/N, La Laguna 38205, Tenerife, Spain}
         \affiliation{
             Departamento de Astrof\'\i sica, Universidad de La Laguna, 38205, Tenerife, Spain}

 \author{Ignacio Trujillo}

        \affiliation{Instituto de Astrof\'\i sica de Canarias, V\'\i a L\'actea S/N, La Laguna 38205, Tenerife, Spain}
         \affiliation{
             Departamento de Astrof\'\i sica, Universidad de La Laguna, 38205, Tenerife, Spain}
\
   \email{hsocas@iac.es, itc@iac.es}
        

 
  \begin{abstract}
The hypothesized Planet Nine is thought to reside in the distant outer solar system, potentially explaining various anomalies in the orbits of extreme trans-Neptunian objects (ETNOs). In this work, we present a targeted observational search for Planet Nine in a field of approximately 98~square-degrees. This field is close to the highest probability region of finding Planet Nine, according to simulations, but poorly constrained by previous searches. Our observations and search methodology, based on the detection of parallax position shifts between consecutive nights, work well in these conditions. We provide 85\% confidence exclusion limits for objects with Sloan $r$-band magnitudes brighter than between 21.0 and 21.4, with an average sensitivity limit of 21.3. No credible Planet Nine candidates were identified within this field and magnitude limits. A caveat to our approach is that it would miss a candidate if its position were affected by scattered light from bright stars in at least one of the nights. However, we estimate that the probability for this is very low, around 0.4\%. We discuss several possible reasons for our Planet Nine non-detection. Our study complements prior searches, particularly those using archival survey data that are limited in the Galactic plane or at fainter brightness limits. While our consecutive-night observation approach offers high sensitivity to minimal motion, extending the search for Planet Nine to fainter magnitudes (which may be crucial, according to recent predictions), will require higher sensitivity instrumentation. 
\end{abstract}

   \maketitle
%

\section{Introduction}

Over the last decades, the exploration of the outer solar system has produced a wealth of new observations, often surprising, that have reshaped our understanding of the solar system's origin and evolution \citep{gladman2021transneptunian}. Many trans-Neptunian objects (TNOs) have been discovered and characterized, revealing a vast and diverse population of objects that helped us understand the early stages of the solar system and its planetary migrations (e.g. \citealt{tsiganis2005origin,levison2008origin,walsh2011low,raymond2014grand}), but also provides intriguing clues that our current picture is not complete. Something is missing in our understanding of the outer solar system, possibly a still undiscovered large and very distant planet (at least). 

The modern Planet Nine hypothesis, not to be confused with earlier conjectures of new solar system planets, has its roots in the discovery of an extreme TNO population (ETNOs) nearly twenty years ago (\citealt{gladman2002evidence}; \citealt{brown2004discovery}). This population is characterized by very large semi-major axis (typically $a \gtrsim$~150~AU) and perihelia further than the orbit of Neptune ($q \gtrsim$~35~AU), which makes them gravitationally detached from the planets. \cite{trujillo2014sedna} noted a statistically unlikely clustering of orbital parameters in the population of known ETNOs and considered the possibility that a massive "unseen perturber" was responsible for the anomaly. The ETNO orbital clustering was further understood by \cite{batygin2016evidence}, who recognized that the orbits were geometrically aligned in physical space and performed a large number of detailed simulations to constrain the orbital parameters of the hypothetical Planet Nine (e.g., \citealt{brown2016observational,khain2018generation,brown2021orbit}). A recent improvement to this approach by \cite{siraj2024orbit} selects a sample of ETNOs based on their secular stability and finds a statistically significant apsidal alignment (clustering in longitude of perihelion) at the 3-$\sigma$ level in the orbits of the 51 stable or metastable ETNOs currently known.

It should be noted that the ETNO orbital alignment is not undisputed, as some authors have pointed out that the apparent parameter clustering could result from observational biases inherent in the surveys that discovered these objects. Given the vast distances and dimness of ETNOs (discovered objects have typical magnitudes between 22 and 24), telescopic surveys are limited in sky coverage and sensitivity, potentially leading to a skewed sample of observed objects. The uneven distribution of telescope locations on Earth, combined with the preferences for specific observing conditions or survey design, may preferentially detect objects in certain areas of the sky, mimicking a clustering effect that does not exist in reality (\citealt{shankman2017ossos,NGD21}). This possibility has been considered by \cite{brown2021orbit} (hereafter BB21) and \cite{siraj2024orbit} who concluded that observational biases are unlikely to explain away the clustering.

In addition to the ETNO orbital alignment, there are other lines of evidence that suggest the existence of this planet. \cite{dLdlFMdlFM17} studied the pair of ETNOs (474640) 2004 VN$_{112}$ (also known as Alicanto) and 2013 RF$_{98}$, concluding that they were most likely a binary asteroid that was detached by a gravitational encounter near their aphelion at some point in solar system history. \cite{dlFM2022nodaldistances} find a very large asymmetry between the distributions of ascending and descending nodal distances of ETNOs, pointing to a secular shepherding of these objects. A recent work by \cite{batygin2024generation} shows that the observed population of TNOs that cross the orbit of Neptune is unstable on timescales of $\sim$100~Myr and must be continuously replenished, most likely by a massive body beyond Neptune. 

Several attempts have been made to search for this unseen planet, the most recent being \cite{BMB24}. See Section~\ref{sec:comparison} for others. Thus far, the efforts have been unfruitful, which has prompted some authors to invoke more exotic sources for the gravitational anomaly in the outer solar system, such as a primordial black hole (\citealt{ScholtzUnwindBHPlanet9}) or MOND effects (\citealt{brown2023MONDPlanet9}). However, a simpler explanation for the lack of success might be the faintness of Planet Nine combined with the large search area, which is basically a fraction of the entire sky. Simulations, such as those performed by BB21 and \cite{siraj2024orbit} are very effective in constraining the orbital parameters of Planet Nine but there is no way to determine its true anomaly, i.e. its current position along the orbit. 

Our search strategy links observations obtained on consecutive or nearby nights to identify sources whose apparent motion is compatible in both direction and amplitude with the predicted parallax of Planet Nine. This represents a novel methodology within Planet Nine searches and allowed us to survey a field located near the region of highest detection probability inferred from dynamical simulations, yet only weakly constrained by previous efforts due to severe source crowding near the Galactic plane.

The present paper is organized as follows: Section~\ref{sec:strategy} describes our observations and the strategy used to identify potential candidates. In Section~\ref{sec:candidates} we discuss the problems and false positives encountered with this strategy. Section~\ref{sec:exclusion} presents the resulting exclusion limits reached in our work. These results are put into the broader context of previous work in Section~\ref{sec:comparison}. Finally, we present our conclusions in Section~\ref{sec:conclusions}.

\section{Observations and search strategy}
\label{sec:strategy}

For our search we rely on the predictions by BB21, who considered various orbital parameters and albedo assumptions, resulting in a likely Sloan $r$-band magnitude between 18 and 22 (throughout this paper, we adopt the AB magnitude system and filters consistent with the Sloan Digital Sky Survey photometric system, see \citealt{Oke74_ABsystem}). It should be noted that these authors recently revised their estimates to a more pessimistic range (\citealt{BMB24}), with a $V$-band magnitude between 20.6 and 23.1. Since this update was not available when our observations were designed, our analysis is based on the original parameter space, with further discussion on this adjustment provided below. The search field is given by SN23, which covers approximately 98 square degrees (17.9\degree$\times$5.5\degree) and contains roughly one million sources with magnitudes between $r$ = 18 and 22. 

Given the extensive number of sources (on the order of 10$^6$), conducting a manual search for an undetected planet is impractical. Historically, solar system planets have been identified through their motion relative to background stars. For a distant planet at hundreds of AU, proper motion is minimal, but parallax may still be detectable. Based on the anticipated range for Planet Nine, parallax should be~20 to~30 times larger than proper motion, resulting in a nightly displacement between 4\arcsec , and 7\arcsec , under optimal observation conditions (Planet Nine at opposition).

Our search strategy is designed to detect a source that shifts by an amount and in a direction consistent with Planet Nine over two consecutive or closely spaced nights. Given the specific sky location, we can calculate expected displacements precisely. The distance remains somewhat uncertain (we rely again on BB21’s predictions), providing a range for potential parallax values. However, the displacement direction is calculated with high accuracy, being determined solely by Earth's orbital motion around the Sun.

For this study, we used data collected with the JAST/T80 telescope and the T80Cam camera (\citealt{T80cam}) at the Observatorio Astrofísico de Javalambre (OAJ) in Teruel. 
The image reduction process was performed using a custom-built pipeline, \texttt{jype}, developed by CEFCA for OAJ surveys (see \citealt{cenarro19}). This pipeline, primarily written in \texttt{python}, employs \texttt{SExtractor} for source extraction and initial photometry and has been under continuous development since 2010. The latest version was used to process the data presented here. In addition to standard reduction steps (bias, flat-field, and fringing corrections when needed), it includes an essential illumination correction required for large-field systems like JAST80. Specifically, standard flat-field corrections on large-field telescopes with field correctors introduce a two-dimensional photometric bias across the images of several tens of mmag, necessitating an additional correction step (see Appendix B.1 of \citealt{bonoli21}). The pipeline also applies aperture photometry corrections within a 6\arcsec \, diameter integration area to estimate the total flux for detected sources.

Observations were conducted over two campaigns: one in 2022 (September 27 and October 2) and the other in 2023 (September 24 and 25). The first campaign encountered suboptimal conditions; although the first night met the seeing requirements (around 1\arcsec), the second night had degraded seeing between 1.5\arcsec \  and 2\arcsec. Since our detection strategy relies on consistent visibility across both nights, the poorer seeing set a limiting magnitude for our search at $r$ = 20.5. Additionally, the time gap between these two nights increased the predicted parallax range to between half and one arc-minute, complicating the pairing of sources between nights and raising the rate of false positives (discussed in Section~\ref{sec:candidates} below). Consequently, this initial search was necessarily modest in scope and results but served as a valuable methodological proof of concept, allowing us to refine our software tools and, more importantly, providing verification data, as discussed below.

The 2023 campaign met all observational criteria, with two full observations of the entire field acquired on two consecutive nights (September 24 and 25). Other attempts yielded incomplete datasets due to insufficient continuous observing conditions. The data from the first campaign and incomplete sets from other nights in the second campaign proved invaluably useful in discarding false positives, which might have otherwise been impossible to rule out. The high false positive rate results from the vast number of sources. We found that, with only two images of the same field, it is not possible to eliminate all image artifacts resembling Planet Nine. Examples are discussed in Section~\ref{sec:candidates}.

\begin{sidewaysfigure*} 
    \centering
    \includegraphics[width=\paperheight]{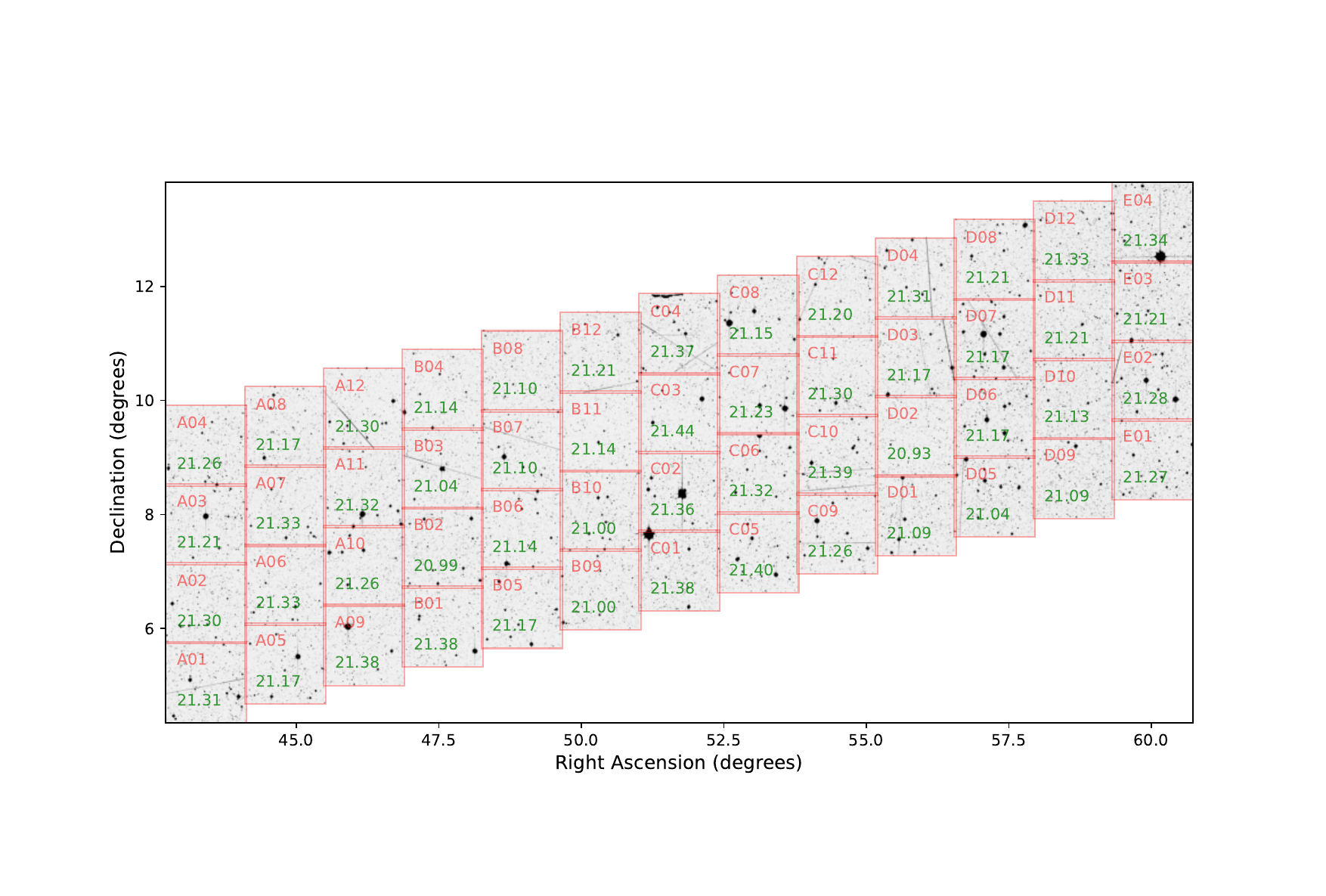}  
    \caption{Mosaic of the observed field. The greyscale represents a negative image with the observed flux on a logarithmic scale to enhance dynamic range. For each subfield we show the best-seeing image. Red: Subfield labels. Green: Sloan $r$-magnitude for which there is an 85\% probability that Planet Nine would have been detected if present.}
    \label{fig:figsky}
\end{sidewaysfigure*}

A mosaic of the observed field is represented in Fig.~\ref{fig:figsky}. Each square is a subfield, identified by a label consisting of a letter (A, B, C, D or E) followed by a two-digit number from 1 to 12. The subfields have the size of the telescope field-of-view and are observed in individual exposures of the T80Cam camera in the SLOAN $r$-filter, having an exposure time of 180~seconds. Each image has a definition of 9200$\times$9200 pixels and covers an area of 1.4\degree$\times$1.4\degree \, on the sky. The figure shows a mosaic combining the best-seeing image acquired for each subfield. The green number is an $r$-magnitude exclusion limit at 85\% confidence, calculated as explained in Section~\ref{sec:exclusion} below.

\section{False positives}
\label{sec:candidates}

We developed an automated search algorithm to parse the T80Cam pipeline’s catalog of sources, aiming to identify pairs of sources that might be the same but shifted between both nights by the expected distance in the direction predicted by our parallax calculations. Conservative tolerances were applied, following the principle that it is less harmful to increase the number of false positives than the risk of excluding Planet Nine. 

Our method begins by generating a list of orphaned sources, defined as those appearing in the catalog from one night but not the other. To identify these, we use the \verb|tskymatch2| tool from the STILTS package (\citealt{STILTS}), comparing sources from the first night against those from the second, and viceversa. The lists of orphaned sources in each night are matched to extract pairs consistent with a displacement in (RA, dec) between (-2.05\arcsec,-0.91\arcsec) and (-8.25\arcsec, -3.69\arcsec). These values define the parallax range for a source at distances between 270 and 1100~AU. We also apply a restriction on the calibrated magnitudes $r$ of each pair, requiring that they differ by no more than 1 magnitude.

This selection process yields a list of 939 candidate pairs. Although inspecting each candidate manually is labor-intensive, it remains feasible. Further reduction of the list could be achieved by applying additional criteria, such as ellipticity or automated cosmic ray detection. However, we opted for a more conservative approach, retaining all identified candidates for manual review.

We then use Skybot (\citealt{skybot}) to compile data on known minor solar system bodies within our field during the observation period. Rather than discarding candidates based on this information, we overlay the positions of known objects on the images for reference during visual inspection.

\begin{figure*} 
    \centering
    \includegraphics[width=0.64\textwidth]{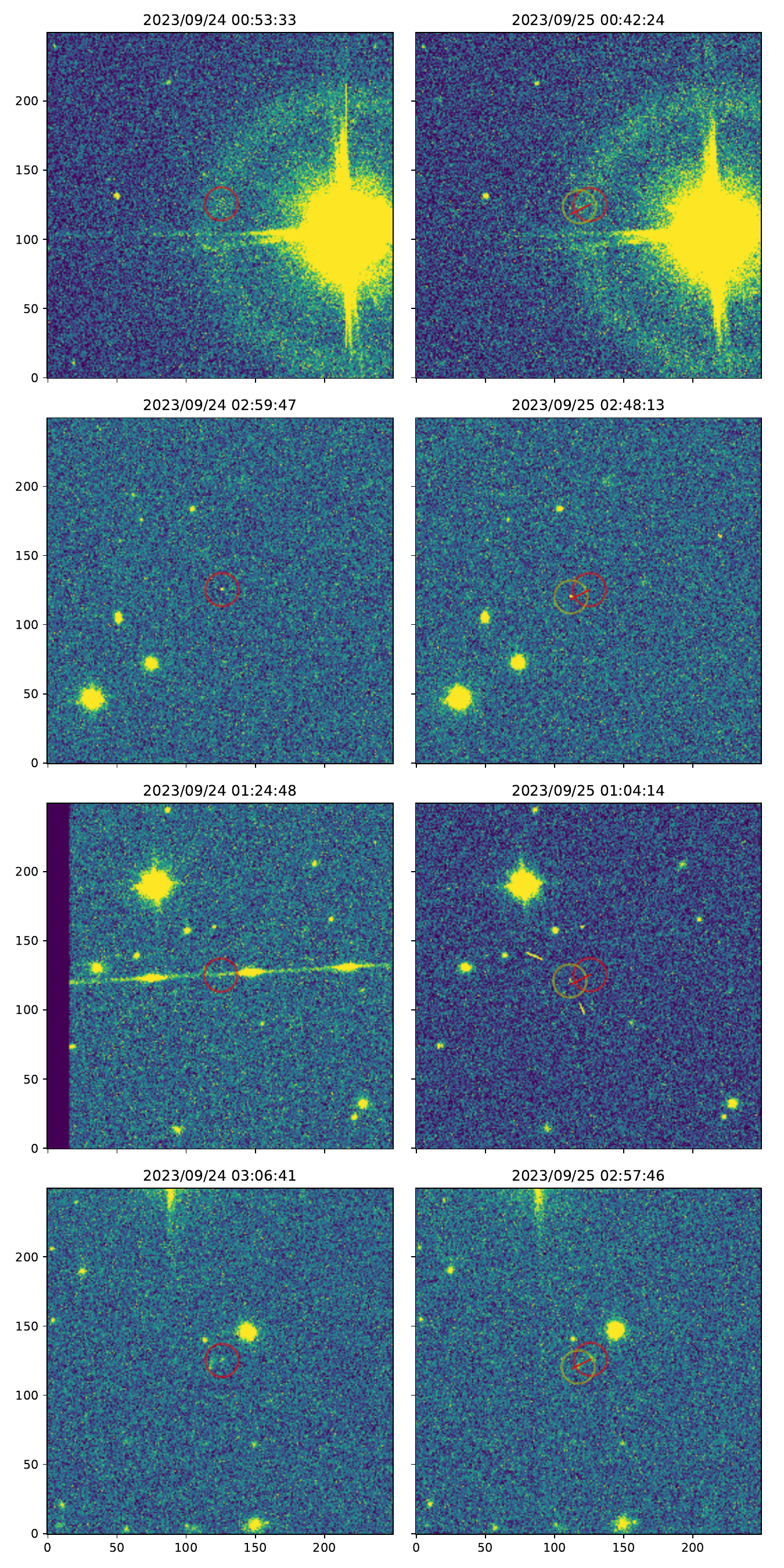}  
    \caption{Four examples of false positives (see text for explanation). Left column: First night. Right column: Second night. The red (yellow) circle marks the source position in the first (second) night, with the red line indicating the apparent motion expected for Planet Nine due to parallax. The $x$ and $y$ units are pixels.}
    \label{fig:candidates}
\end{figure*}

Figure~\ref{fig:candidates}
provides examples of common false positives in our methodology. The most frequent artifacts arise from diffraction spikes or halos around bright stars, as shown in the top row. These brightness fluctuations can sometimes be misidentified as sources by the algorithm. A small subset of these spurious sources occurs in pairs that meet our search criteria, thus being flagged as candidates. Although the probability of such coincidences is low, the large number of sources we analyze makes this type of false positive prevalent, dominating the 939 candidates we examined manually.

Cosmic rays are the second largest source of false positives. In some cases, a cosmic ray hits during the first night, and on the second night another cosmic ray coincidentally appears close to where Planet Nine would have been if it had been the object observed in the first night. An example of this situation is shown in the second row of Fig.~\ref{fig:candidates}. While this coincidence might seem unlikely, the extensive field area made this scenario a rather common occurrence. For some candidates, such as the one shown, ruling them out would have been impossible without the additional images from the extra dataset or the first campaign.

Sometimes, different confounding factors combine to produce pairs. In the third row of Fig.\ref{fig:candidates}, a false positive is created by a combination of an artificial satellite on the first night and a cosmic ray on the second. Another notable source of false positives is the intermittent detection of the faintest sources, those near the noise threshold. The source identification algorithm may occasionally miss these faint sources, leading to instances where only one of the pair is detected each night. An example of this situation is shown in the fourth row of Fig.\ref{fig:candidates}.

After careful manual inspection, we ruled out all 939 candidates as potential Planet Nine detections. In some cases, additional observations were necessary, which is a valuable lesson for future searches of a similar nature. Highly improbable combinations of artifacts do occur given the large number of sources and the expansive sky area under investigation.

\section{Exclusion limits}
\label{sec:exclusion}

To convert our non-detection of Planet Nine into meaningful scientific constraints, we need to establish the limiting magnitudes derived from our search. In other words, it is necessary to estimate the range of the parameter space that our observations rule out. A simple approach would be to determine the limiting magnitude of our data and conclude that objects brighter than this threshold are excluded.

However, this approach would be overly simplistic, as there is no sharp magnitude threshold for our search. A faint source with flux near the noise level has a variable probability of detection, depending on the specific noise realization or the presence of image artifacts.

\begin{figure*}[p]  
    \centering
    \includegraphics[width=1.8\textwidth,height=\textheight,keepaspectratio,trim=0 0 0 -20, clip]{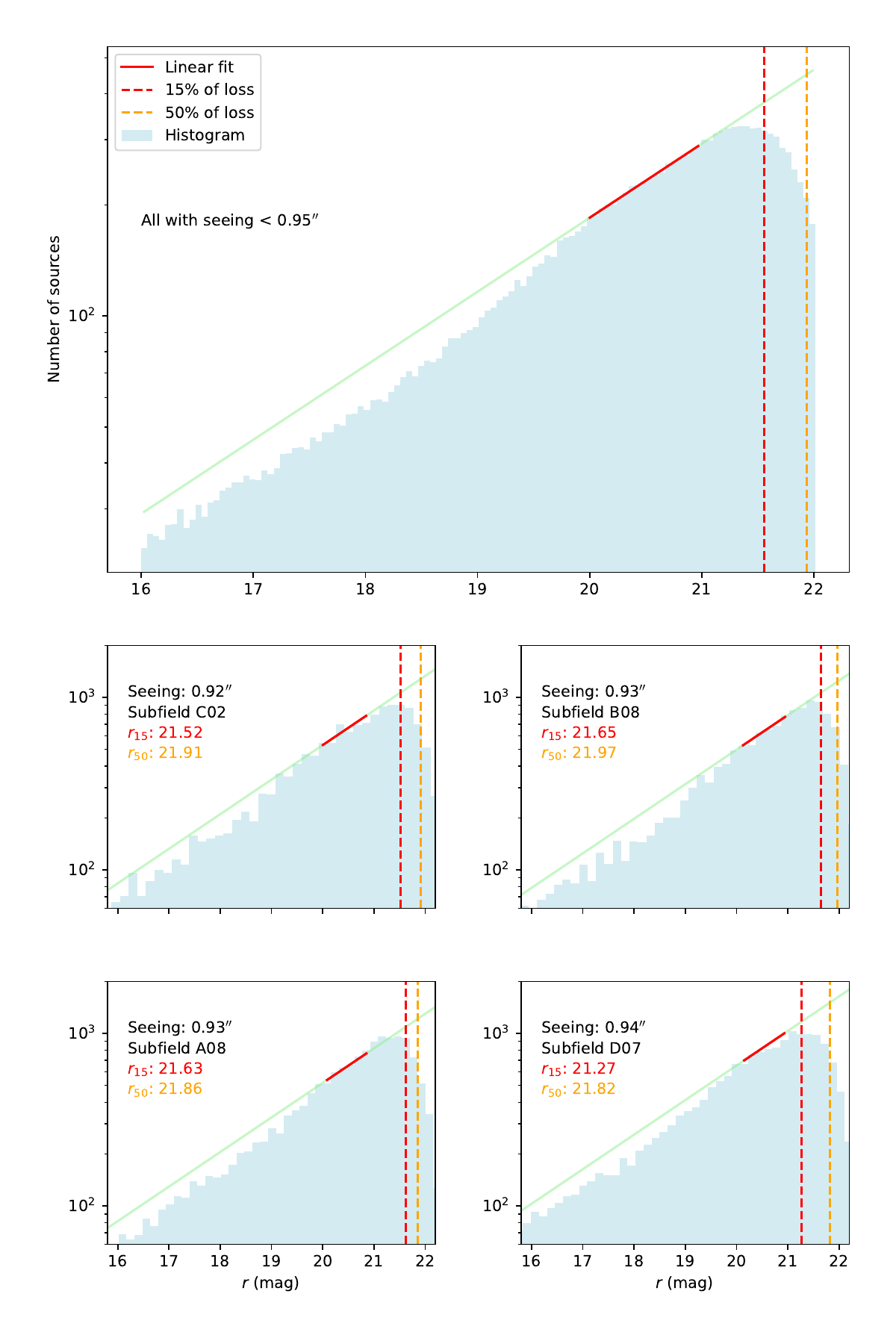}
    \caption{Histograms of the number of sources detected per magnitude interval.
      Green line: Linear fit. Red line: Range of data fitted. Vertical dashed lines: Magnitudes at which 85\% (red) and 50\% (yellow) of the sources are detected (assuming that the true number of sources is well approximated by the fit, see text).
      Upper panel: Combining all frames with seeing better than 0.95\arcsec .
      Lower panels: Histograms for a sample of randomly selected subfields. 
    }
    \label{fig:histograms_fits}
\end{figure*}

In general, the number of sources detected in an image increases with magnitude, approximately following a power law. Figure~\ref{fig:histograms_fits}
(top panel) illustrates this trend in our highest-quality images (those taken with seeing better than 0.95\arcsec). This threshold balances image quality with the number of frames available, resulting in 27 selected frames. Selecting an optimal bin size involves a trade-off between the resolution of the resulting curve and statistical fluctuations due to bin sample size. We used the Freedman-Diaconis rule (\citealt{freedmandiaconis}) to determine the optimal bin width, which in this case is 0.054 magnitudes. The number of sources shows a power-law increase up to the sensitivity limit, beyond which it begins to decline.

Assuming that this drop-off results from our reduced sensitivity to faint sources and that the true source distribution remains the same power law near the elbow, we estimate the number of sources we are likely missing. By comparing our observed sources to an extrapolated distribution (which we assume accurately models the true source population), we determine the magnitude at which we start missing a certain percentage of sources. We consider 15\% and 50\% as meaningful thresholds and report our results accordingly, as 85\% and 50\% confidence. It is worth mentioning that this empirical method combines both point-like and extended sources. In this sense, we expect the magnitudes provided here to be slightly below the ones expected for point-like sources and therefore to be conservative.

The four lower panels in Figure~\ref{fig:histograms_fits} display similar histograms for individual images, which represent a subset of our best observations (four images randomly selected with seeing better than 0.95\arcsec). These individual histograms are noisier than the composite in the top panel, which combines data from 27 images. The optimal Freedman-Diaconis bin width here is 0.15 magnitudes. The linear fit from the combined histogram (the one shown in the top panel) is overlaid on each individual frame histogram, indicating that the same power-law relationship between source count and magnitude holds across all images. Minor differences between frames arise from variations in the number of sources, which slightly shifts the curve across exposures. We adjust for these differences by normalizing the linear model to the number of sources in the range of $r$ between~20 and~20.5.

We define the parameters $r_{15}$ and $r_{50}$ as the magnitudes at which we miss 15\% and 50\% of existing sources in a given image. These values can also be interpreted as the magnitudes for which we have a 15\% or 50\% probability of missing a source actually present in the image. Generally, the fraction of missed sources $m(r)$ is a distribution defined by:
\begin{equation}
  m(r)= \frac {  N_{True}(r) - N_{Obs}(r)}{N_{True}(r)} \, ,
\end{equation}
where $N_{True}(r)$ and $N_{Obs}(r)$ are the distributions of true and observed sources, i.e.: $N(r) dr$ is the number of sources with magnitudes in the interval from $r$ to $r + dr$. The true number of sources is assumed to be well represented by the linear fit and, since $0 \le N_{Obs}(r) \le N_{True}(r)$, $m(r)$ ranges from $0$ to $1$.

\begin{figure}[ht]  
    \centering
    \includegraphics[width=0.8\textwidth]{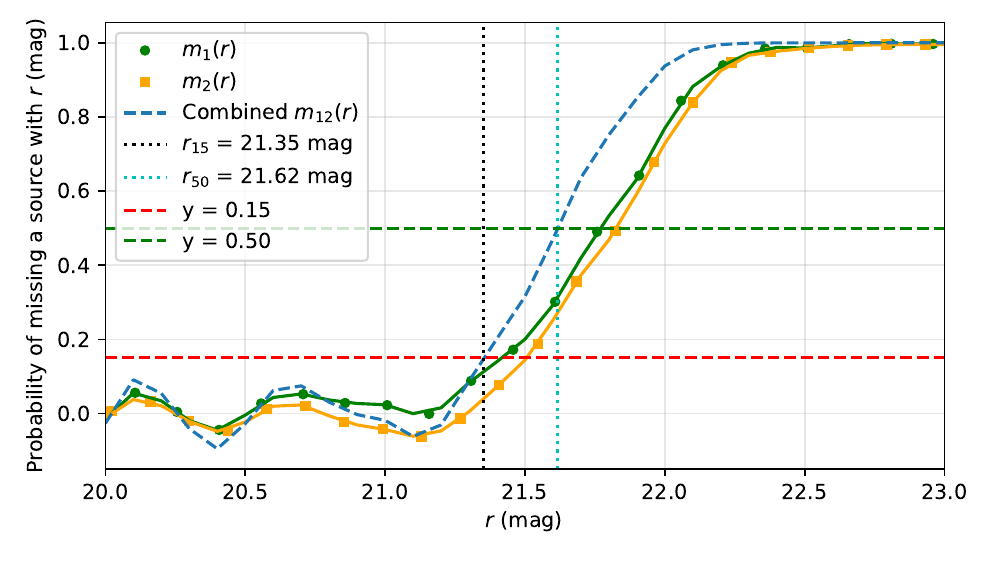}
    \caption{Estimated probability of missing sources ($m_1(r)$ and $m_2(r)$) and combined probability $m_{12}(r)$ of not detecting Planet Nine in the two best images of subfield A01. Symbols show original histogram values, with curves interpolated to a finer grid using cubic splines. Negative values of $m$ on the left side of the curve are spurious, caused by random fluctuations around zero as the number of observed sources $N_{Obs}$ fluctuates above and below the fit that approximates the true number of sources $N_{True}$. This regime is irrelevant for our analysis.}
    \label{fig:prob_combined}
\end{figure}

Detecting Planet Nine with our approach requires successful detection in two separate images of the same field. Since each detection is independent, the combined probability of missing the object, $m_{12}(r)$, can be calculated from the probabilities of missing it in either frame, $m_1(r)$ or $m_2(r)$:
\begin{equation}
    1-m_{12}(r)=(1-m_1(r))(1-m_2(r)) \, .
\end{equation}
This combined probability is illustrated in Figure~\ref{fig:prob_combined}, showing the curves for the two best images of subfield A01. Naturally, the probability of missing Planet Nine at a given magnitude is higher when considering both frames than for either frame alone, as detection in both images is required for a successful identification.

Using this approach, we calculate the $r_{15}$ and $r_{50}$ values for each image pair in each subfield, selecting the best pair to represent the exclusion limits of our search for that subfield. Table~\ref{table:tab1} presents these results (see also Fig~\ref{fig:figsky}).

\begin{table}
\label{table:tab1}
\caption{Limiting magnitudes $r_{15}$ and $r_{50}$ in each subfield}
\centering
\begin{tabular}{lcc|lcc}
\hline\hline
 & $r_{{15}}$ & $r_{{50}}$ &  & $r_{{15}}$ & $r_{{50}}$ \\
Subfield & (mag) & (mag) & Subfield & (mag) & (mag) \\
\hline
A01 & 21.31 & 21.59 & C03 & 21.44 & 21.64 \\
A02 & 21.30 & 21.59 & C04 & 21.37 & 21.71 \\
A03 & 21.21 & 21.57 & C05 & 21.40 & 21.68 \\
A04 & 21.26 & 21.62 & C06 & 21.32 & 21.68 \\
A05 & 21.17 & 21.56 & C07 & 21.23 & 21.67 \\
A06 & 21.33 & 21.64 & C08 & 21.15 & 21.59 \\
A07 & 21.33 & 21.62 & C09 & 21.26 & 21.67 \\
A08 & 21.17 & 21.55 & C10 & 21.39 & 21.68 \\
A09 & 21.38 & 21.62 & C11 & 21.30 & 21.62 \\
A10 & 21.26 & 21.63 & C12 & 21.20 & 21.59 \\
A11 & 21.32 & 21.63 & D01 & 21.09 & 21.51 \\
A12 & 21.30 & 21.61 & D02 & 20.93 & 21.50 \\
B01 & 21.38 & 21.64 & D03 & 21.17 & 21.55 \\
B02 & 20.99 & 21.49 & D04 & 21.31 & 21.58 \\
B03 & 21.04 & 21.50 & D05 & 21.04 & 21.49 \\
B04 & 21.14 & 21.51 & D06 & 21.17 & 21.56 \\
B05 & 21.17 & 21.48 & D07 & 21.17 & 21.56 \\
B06 & 21.14 & 21.52 & D08 & 21.21 & 21.56 \\
B07 & 21.10 & 21.48 & D09 & 21.09 & 21.55 \\
B08 & 21.10 & 21.55 & D10 & 21.13 & 21.55 \\
B09 & 21.00 & 21.52 & D11 & 21.21 & 21.62 \\
B10 & 21.00 & 21.43 & D12 & 21.33 & 21.61 \\
B11 & 21.14 & 21.45 & E01 & 21.27 & 21.55 \\
B12 & 21.21 & 21.52 & E02 & 21.28 & 21.60 \\
C01 & 21.38 & 21.69 & E03 & 21.21 & 21.56 \\
C02 & 21.36 & 21.67 & E04 & 21.34 & 21.67 \\
\hline
\end{tabular}
\end{table}

\section{Comparison with previous work}
\label{sec:comparison}

The most recent large-scale search for Planet Nine was conducted by \cite{BMB24}, using a compilation of data from several wide-field surveys, including Pan-STARRS1, the Zwicky Transient Facility (ZTF), and the Dark Energy Survey (DES), to examine the sky regions predicted by Planet Nine simulations (\citealt{BB21}). Unlike our approach, these surveys were not specifically tailored for Planet Nine detection; rather, they leveraged extensive archival data with observations spanning multiple years. For instance, the Pan-STARRS1 data alone covered a five-year period, over an extensive area of the sky.

From this vast dataset, over a billion individual detections were initially flagged. To make the data manageable, a series of selection filters was applied. A key filtering step involved applying a mask to exclude data from regions of the detector deemed unreliable. Like our methodology, they identified bright star surroundings as problematic, which in their case required these regions to be masked. Ultimately, the number of viable detections was narrowed to approximately 244 million. Each of these detections was then cross-matched to identify whether any set could represent a solar system object moving on a Keplerian orbit. To ensure reliability, they required a minimum of nine consistent detections per object to confirm potential orbital alignment, given that even a threshold of seven detections resulted in an unmanageable volume of candidates.

To evaluate the robustness of their method, \cite{BMB24} introduced a population of synthetic objects with characteristics predicted for Planet Nine, constructing a map of exclusion limits based on the detectability of these synthetic objects. Their results offered comprehensive sky coverage, with some limitations. Regions where Planet Nine's hypothesized orbit intersects with the Galactic plane and adjacent areas have lower coverage due to high source density and the fact that surveys are often optimized to avoid these complex regions when they are designed for other scientific objectives. Thus, certain portions of the sky, particularly those near the Galaxy, remain less constrained (see their Figure~4).

Our study complements the work of \cite{BMB24} by targeting a distinct and under-explored region of the sky. We focused on an area that serendipitously coincides with a region where \cite{BMB24} have a lower coverage. Specifically, their synthetic population model estimated that the likelihood of missing Planet Nine in our search field ranged from approximately 8\% to 50\%, depending on the subfield, due to limited coverage. Therefore, our work provides refined exclusion limits in a difficult observational region.

The TESS exoplanet satellite is also being used to search for Planet Nine. \cite{rice2020} employed a novel approach utilizing full-frame images from TESS, focusing on a technique called shift-stacking to enhance signal detection in a dense stellar environment, specifically along the Galactic plane. Their method involved co-adding images aligned along hypothetical orbital paths, thereby summing small flux increments across frames to potentially reveal faint, slow-moving solar system objects that might otherwise be lost in the noise. Their approach would allow to detect a Planet Nine brighter than magnitude $V$ = 21 and closer than 150~AU. 

One of the most extensive searches for Planet Nine using infrared data was conducted by \cite{Meisner2017} and extended in \cite{Meisner2018}, where they utilized a customized analysis of WISE and NEOWISE data to examine a substantial portion of the sky. Their search methodology focused on co-adding W1 (3.4 $\mu$m) exposures to increase sensitivity, allowing them to detect much fainter objects than could be achieved with single WISE frames alone. They primarily targeted the high Galactic latitude sky regions, covering roughly three-quarters of the sky with a magnitude limit of $W1$ < 16.7 at 90\% completeness.

In contrast to our targeted, short-timescale observation strategy, \cite{Meisner2017} and \cite{Meisner2018} leveraged archival data over a seven-year baseline. This allowed them to cover large sky areas but required complex processing to handle the extensive temporal spacing between frames, with coadds spanning several years. This approach is ideal for capturing faint sources moving along a Keplerian orbit over time, particularly for distant or slowly moving objects, but it is less sensitive to small, consecutive-night displacements detectable in our strategy and, crucially, is limited at low Galactic latitudes, where our search is focused. They reported limitations in regions near the Galactic plane due to high source density and increased noise, which complicated the differentiation of faint objects from background sources.

\section{Conclusions}
\label{sec:conclusions}

Our search did not reveal any source that could be considered a credible Planet Nine candidate. This non-detection may stem from one or more of the following reasons:

\begin{itemize}

\item Planet Nine does not exist. Until direct evidence of such a body is obtained, this possibility remains. Historically, some planets were successfully predicted through their gravitational effects on other objects, yet others were hypothesized due to perceived anomalies that later proved spurious. While caution is warranted, the Planet Nine hypothesis is supported by several independent lines of evidence. Even if the clustering of extreme trans-Neptunian object (ETNO) orbits were due to observational bias, there would still be other observations needing alternative explanations, including the Neptune-crossing TNO population (\citealt{batygin2024generation}), the intriguing orbital detachment of the ETNOs Alicanto and 2013~RF$_{98}$ (\citealt{dLdlFMdlFM17}), and the asymmetry in the ascending and descending nodal distances of known TNOs (\citealt{dlFM2022nodaldistances}).

\item Perhaps Planet Nine exists but it is not located where we searched. Our search strategy is based on the hypothesis of SN23, which proposes that the trajectory of the peculiar CNEOS14 bolide was altered by an interaction with an unknown planet in the outer solar system. This scenario is controversial, not only by its own assumptions but also because of doubts on the CNEOS database, as technical details about its detectors are classified. Although some bolide events have been simultaneously measured by ground-based scientific observatories, allowing for detailed cross-calibration, there remains considerable debate. For example, \cite{bb23} argue that the interstellar nature of CNEOS14 is compromised by an alleged correlation between measurement errors and event speed, a claim later disputed by \cite{sn24}. A recent paper by \cite{PASNS25} suggests that CNEOS14 might belong to the population of events measured with large errors in the CNEOS database. The interstellar nature of this meteoroid is at this point a claim under dispute.

\item Image artifacts. One of the images of Planet Nine might have been obscured in our observations by a nearby bright star or another image artifact, preventing its detection. However, this would be very unlikely since the fraction of pixels with brightness significantly above the background noise in our images is $\simeq$0.4\%.

\item Planet Nine may be fainter than our detection limits. Our search assumed an optimistic brightness estimate for Planet Nine, between $r$-band magnitudes~18 and~22. covering roughly the 84th percentile of brightness predicted by \cite{BB21}. Our 85\% confidence exclusion limits range between magnitudes 20.7 and 21.5 across different subfields, with an average limit of 21.3. Considering both our detection confidence and the 84th percentile assumption, there remains about a 30\% probability that we would miss Planet Nine if it falls within this brightness range. A more recent work by \cite{BMB24} provides a revised $V$-band magnitude estimate between~20.6 and~23.1, much of which lies beyond our sensitivity and would require a larger telescope and/or longer integration times.

\end{itemize}

Given the distinct possibility that Planet Nine may fall outside the sensitivity limits of our observations, it is worthwhile to continue this search effort with instruments offering higher sensitivity. Our work complements previous surveys, providing additional constraints within a specific field and reaching depth limits not covered by some of the broader archival and survey data. As such, this study serves as another step in narrowing down Planet Nine’s potential location and demonstrates a methodology that is relatively simple and therefore robust to degeneracies and other common problems.

\begin{acknowledgements}
Based on observations made with the JAST80 telescope at the
Observatorio Astrofísico de Javalambre (OAJ), in Teruel, owned,
managed and operated by the Centro de Estudios de Física del Cosmos de
Aragón (CEFCA). We are grateful to the CEFCA for allocation of
Director’s Discretionary Time to this program. We thank the OAJ Data
Processing and Archiving Unit (UPAD) for reducing and calibrating the
OAJ data used in this work. IT acknowledges support from the ACIISI, Consejer\'{i}a de
Econom\'{i}a, Conocimiento y Empleo del Gobierno de Canarias and the
European Regional Development Fund (ERDF) under a grant with reference
PROID2021010044 and from the State Research Agency (AEI-MCINN) of the
Spanish Ministry of Science and Innovation under the grant
PID2022-140869NB-I00 and IAC project P/302302, financed by the Ministry
of Science and Innovation, through the State Budget and by the Canary
Islands Department of Economy, Knowledge, and Employment, through the
Regional Budget of the Autonomous Community.
This research utilized NASA's Astrophysics Data System Bibliographic Services, alongside the CNEOS and Horizons databases from the Jet Propulsion Laboratory. The figures and data analyses in this paper were generated with Python modules including Matplotlib \citep{H07}, Numpy \citep{numpy11}, and iPython \citep{ipython07}.

\end{acknowledgements}

%
%
\bibliography{paper.bib} 

@inproceedings{STILTS,
  title={Stilts-a package for command-line processing of tabular data},
  author={Taylor, Mark B},
  booktitle={Astronomical Data Analysis Software and Systems XV},
  volume={351},
  pages={666},
  year={2006}
}

@ARTICLE{Oke74_ABsystem,
       author = {{Oke}, J.~B.},
        title = "{Absolute Spectral Energy Distributions for White Dwarfs}",
      journal = {\apjs},
         year = 1974,
        month = feb,
       volume = {27},
        pages = {21},
          doi = {10.1086/190287},
       adsurl = {https://ui.adsabs.harvard.edu/abs/1974ApJS...27...21O}
}

@INPROCEEDINGS{skybot,
   author = {{Berthier}, J. and {Vachier}, F. and {Thuillot}, W. and {Fernique}, P. and 
             {Ochsenbein}, F. and {Genova}, F. and {Lainey}, V. and {Arlot}, J.-E.},
    title = "{SkyBoT, a new VO service to identify Solar System objects}",
booktitle = {Astronomical Data Analysis Software and Systems XV},
     year = 2006,
   series = {Astronomical Society of the Pacific Conference Series},
   volume = 351,
   editor = {{Gabriel}, C. and {Arviset}, C. and {Ponz}, D. and {Enrique}, S.},
    month = jul,
    pages = {367-+},
   adsurl = {http://adsabs.harvard.edu/abs/2006ASPC..351..367B}
}

@inproceedings{T80cam,
  title={T80Cam: the wide field camera for the OAJ 83-cm telescope},
  author={Marin-Franch, A and Taylor, K and Cepa, J and Laporte, R and Cenarro, AJ and Chueca, S and Cristobal-Hornillos, D and Ederoclite, A and Gruel, N and Hern{\'a}ndez-Fuertes, J and others},
  booktitle={Ground-based and Airborne Instrumentation for Astronomy IV},
  volume={8446},
  pages={1925--1931},
  year={2012},
  organization={SPIE}
}

@ARTICLE{cenarro19,
       author = {{Cenarro}, A.~J. and {Moles}, M. and {Crist{\'o}bal-Hornillos}, D. and {Mar{\'\i}n-Franch}, A. and {Ederoclite}, A. and {Varela}, J. and {L{\'o}pez-Sanjuan}, C. and {Hern{\'a}ndez-Monteagudo}, C. and {Angulo}, R.~E. and {V{\'a}zquez Rami{\'o}}, H. and {Viironen}, K. and {Bonoli}, S. and {Orsi}, A.~A. and {Hurier}, G. and {San Roman}, I. and {Greisel}, N. and {Vilella-Rojo}, G. and {D{\'\i}az-Garc{\'\i}a}, L.~A. and {Logro{\~n}o-Garc{\'\i}a}, R. and {Gurung-L{\'o}pez}, S. and {Spinoso}, D. and {Izquierdo-Villalba}, D. and {Aguerri}, J.~A.~L. and {Allende Prieto}, C. and {Bonatto}, C. and {Carvano}, J.~M. and {Chies-Santos}, A.~L. and {Daflon}, S. and {Dupke}, R.~A. and {Falc{\'o}n-Barroso}, J. and {Gon{\c{c}}alves}, D.~R. and {Jim{\'e}nez-Teja}, Y. and {Molino}, A. and {Placco}, V.~M. and {Solano}, E. and {Whitten}, D.~D. and {Abril}, J. and {Ant{\'o}n}, J.~L. and {Bello}, R. and {Bielsa de Toledo}, S. and {Castillo-Ram{\'\i}rez}, J. and {Chueca}, S. and {Civera}, T. and {D{\'\i}az-Mart{\'\i}n}, M.~C. and {Dom{\'\i}nguez-Mart{\'\i}nez}, M. and {Garzar{\'a}n-Calderaro}, J. and {Hern{\'a}ndez-Fuertes}, J. and {Iglesias-Marzoa}, R. and {I{\~n}iguez}, C. and {Jim{\'e}nez Ruiz}, J.~M. and {Kruuse}, K. and {Lamadrid}, J.~L. and {Lasso-Cabrera}, N. and {L{\'o}pez-Alegre}, G. and {L{\'o}pez-Sainz}, A. and {Ma{\'\i}cas}, N. and {Moreno-Signes}, A. and {Muniesa}, D.~J. and {Rodr{\'\i}guez-Llano}, S. and {Rueda-Teruel}, F. and {Rueda-Teruel}, S. and {Soriano-Lagu{\'\i}a}, I. and {Tilve}, V. and {Valdivielso}, L. and {Yanes-D{\'\i}az}, A. and {Alcaniz}, J.~S. and {Mendes de Oliveira}, C. and {Sodr{\'e}}, L. and {Coelho}, P. and {Lopes de Oliveira}, R. and {Tamm}, A. and {Xavier}, H.~S. and {Abramo}, L.~R. and {Akras}, S. and {Alfaro}, E.~J. and {Alvarez-Candal}, A. and {Ascaso}, B. and {Beasley}, M.~A. and {Beers}, T.~C. and {Borges Fernandes}, M. and {Bruzual}, G.~R. and {Buzzo}, M.~L. and {Carrasco}, J.~M. and {Cepa}, J. and {Cortesi}, A. and {Costa-Duarte}, M.~V. and {De Pr{\'a}}, M. and {Favole}, G. and {Galarza}, A. and {Galbany}, L. and {Garcia}, K. and {Gonz{\'a}lez Delgado}, R.~M. and {Gonz{\'a}lez-Serrano}, J.~I. and {Guti{\'e}rrez-Soto}, L.~A. and {Hernandez-Jimenez}, J.~A. and {Kanaan}, A. and {Kuncarayakti}, H. and {Landim}, R.~C.~G. and {Laur}, J. and {Licandro}, J. and {Lima Neto}, G.~B. and {Lyman}, J.~D. and {Ma{\'\i}z Apell{\'a}niz}, J. and {Miralda-Escud{\'e}}, J. and {Morate}, D. and {Nogueira-Cavalcante}, J.~P. and {Novais}, P.~M. and {Oncins}, M. and {Oteo}, I. and {Overzier}, R.~A. and {Pereira}, C.~B. and {Rebassa-Mansergas}, A. and {Reis}, R.~R.~R. and {Roig}, F. and {Sako}, M. and {Salvador-Rusi{\~n}ol}, N. and {Sampedro}, L. and {S{\'a}nchez-Bl{\'a}zquez}, P. and {Santos}, W.~A. and {Schmidtobreick}, L. and {Siffert}, B.~B. and {Telles}, E. and {Vilchez}, J.~M.},
        title = "{J-PLUS: The Javalambre Photometric Local Universe Survey}",
      journal = {\aap},
         year = 2019,
        month = feb,
       volume = {622},
          eid = {A176},
        pages = {A176},
          doi = {10.1051/0004-6361/201833036},
archivePrefix = {arXiv},
       eprint = {1804.02667},
 primaryClass = {astro-ph.GA},
       adsurl = {https://ui.adsabs.harvard.edu/abs/2019A&A...622A.176C}
}

@ARTICLE{bonoli21,
       author = {{Bonoli}, S. and {Mar{\'\i}n-Franch}, A. and {Varela}, J. and {V{\'a}zquez Rami{\'o}}, H. and {Abramo}, L.~R. and {Cenarro}, A.~J. and {Dupke}, R.~A. and {V{\'\i}lchez}, J.~M. and {Crist{\'o}bal-Hornillos}, D. and {Gonz{\'a}lez Delgado}, R.~M. and {Hern{\'a}ndez-Monteagudo}, C. and {L{\'o}pez-Sanjuan}, C. and {Muniesa}, D.~J. and {Civera}, T. and {Ederoclite}, A. and {Hern{\'a}n-Caballero}, A. and {Marra}, V. and {Baqui}, P.~O. and {Cortesi}, A. and {Cypriano}, E.~S. and {Daflon}, S. and {de Amorim}, A.~L. and {D{\'\i}az-Garc{\'\i}a}, L.~A. and {Diego}, J.~M. and {Mart{\'\i}nez-Solaeche}, G. and {P{\'e}rez}, E. and {Placco}, V.~M. and {Prada}, F. and {Queiroz}, C. and {Alcaniz}, J. and {Alvarez-Candal}, A. and {Cepa}, J. and {Maroto}, A.~L. and {Roig}, F. and {Siffert}, B.~B. and {Taylor}, K. and {Benitez}, N. and {Moles}, M. and {Sodr{\'e}}, L. and {Carneiro}, S. and {Mendes de Oliveira}, C. and {Abdalla}, E. and {Angulo}, R.~E. and {Aparicio Resco}, M. and {Balaguera-Antol{\'\i}nez}, A. and {Ballesteros}, F.~J. and {Brito-Silva}, D. and {Broadhurst}, T. and {Carrasco}, E.~R. and {Castro}, T. and {Cid Fernandes}, R. and {Coelho}, P. and {de Melo}, R.~B. and {Doubrawa}, L. and {Fernandez-Soto}, A. and {Ferrari}, F. and {Finoguenov}, A. and {Garc{\'\i}a-Benito}, R. and {Iglesias-P{\'a}ramo}, J. and {Jim{\'e}nez-Teja}, Y. and {Kitaura}, F.~S. and {Laur}, J. and {Lopes}, P.~A.~A. and {Lucatelli}, G. and {Mart{\'\i}nez}, V.~J. and {Maturi}, M. and {Overzier}, R.~A. and {Pigozzo}, C. and {Quartin}, M. and {Rodr{\'\i}guez-Mart{\'\i}n}, J.~E. and {Salzano}, V. and {Tamm}, A. and {Tempel}, E. and {Umetsu}, K. and {Valdivielso}, L. and {von Marttens}, R. and {Zitrin}, A. and {D{\'\i}az-Mart{\'\i}n}, M.~C. and {L{\'o}pez-Alegre}, G. and {L{\'o}pez-Sainz}, A. and {Yanes-D{\'\i}az}, A. and {Rueda-Teruel}, F. and {Rueda-Teruel}, S. and {Abril Iba{\~n}ez}, J. and {L Ant{\'o}n Bravo}, J. and {Bello Ferrer}, R. and {Bielsa}, S. and {Casino}, J.~M. and {Castillo}, J. and {Chueca}, S. and {Cuesta}, L. and {Garzar{\'a}n Calderaro}, J. and {Iglesias-Marzoa}, R. and {{\'I}niguez}, C. and {Lamadrid Gutierrez}, J.~L. and {Lopez-Martinez}, F. and {Lozano-P{\'e}rez}, D. and {Ma{\'\i}cas Sacrist{\'a}n}, N. and {Molina-Ib{\'a}{\~n}ez}, E.~L. and {Moreno-Signes}, A. and {Rodr{\'\i}guez Llano}, S. and {Royo Navarro}, M. and {Tilve Rua}, V. and {Andrade}, U. and {Alfaro}, E.~J. and {Akras}, S. and {Arnalte-Mur}, P. and {Ascaso}, B. and {Barbosa}, C.~E. and {Beltr{\'a}n Jim{\'e}nez}, J. and {Benetti}, M. and {Bengaly}, C.~A.~P. and {Bernui}, A. and {Blanco-Pillado}, J.~J. and {Borges Fernandes}, M. and {Bregman}, J.~N. and {Bruzual}, G. and {Calderone}, G. and {Carvano}, J.~M. and {Casarini}, L. and {Chaves-Montero}, J. and {Chies-Santos}, A.~L. and {Coutinho de Carvalho}, G. and {Dimauro}, P. and {Duarte Puertas}, S. and {Figueruelo}, D. and {Gonz{\'a}lez-Serrano}, J.~I. and {Guerrero}, M.~A. and {Gurung-L{\'o}pez}, S. and {Herranz}, D. and {Huertas-Company}, M. and {Irwin}, J.~A. and {Izquierdo-Villalba}, D. and {Kanaan}, A. and {Kehrig}, C. and {Kirkpatrick}, C.~C. and {Lim}, J. and {Lopes}, A.~R. and {Lopes de Oliveira}, R. and {Marcos-Caballero}, A. and {Mart{\'\i}nez-Delgado}, D. and {Mart{\'\i}nez-Gonz{\'a}lez}, E. and {Mart{\'\i}nez-Somonte}, G. and {Oliveira}, N. and {Orsi}, A.~A. and {Penna-Lima}, M. and {Reis}, R.~R.~R. and {Spinoso}, D. and {Tsujikawa}, S. and {Vielva}, P. and {Vitorelli}, A.~Z. and {Xia}, J.~Q. and {Yuan}, H.~B. and {Arroyo-Polonio}, A. and {Dantas}, M.~L.~L. and {Galarza}, C.~A. and {Gon{\c{c}}alves}, D.~R. and {Gon{\c{c}}alves}, R.~S. and {Gonzalez}, J.~E. and {Gonzalez}, A.~H. and {Greisel}, N. and {Jim{\'e}nez-Esteban}, F. and {Landim}, R.~G. and {Lazzaro}, D. and {Magris}, G. and {Monteiro-Oliveira}, R. and {Pereira}, C.~B. and {Rebou{\c{c}}as}, M.~J. and {Rodriguez-Espinosa}, J.~M. and {Santos da Costa}, S. and {Telles}, E.},
        title = "{The miniJPAS survey: A preview of the Universe in 56 colors}",
      journal = {\aap},
         year = 2021,
        month = sep,
       volume = {653},
          eid = {A31},
        pages = {A31},
          doi = {10.1051/0004-6361/202038841},
archivePrefix = {arXiv},
       eprint = {2007.01910},
 primaryClass = {astro-ph.CO},
       adsurl = {https://ui.adsabs.harvard.edu/abs/2021A&A...653A..31B}
}

@article{freedmandiaconis,
  title={On the histogram as a density estimator: L 2 theory},
  author={Freedman, David and Diaconis, Persi},
  journal={Zeitschrift f{\"u}r Wahrscheinlichkeitstheorie und verwandte Gebiete},
  volume={57},
  number={4},
  pages={453--476},
  year={1981},
  publisher={Springer-Verlag Berlin/Heidelberg}
}

@article{brown2016observational,
  title={Observational constraints on the orbit and location of planet nine in the outer solar system},
  author={Brown, Michael E and Batygin, Konstantin},
  journal={The Astrophysical Journal Letters},
  volume={824},
  number={2},
  pages={L23},
  year={2016},
  publisher={IOP Publishing}
}

@article{batygin2024generation,
  title={Generation of Low-inclination, Neptune-crossing Trans-Neptunian Objects by Planet Nine},
  author={Batygin, Konstantin and Morbidelli, Alessandro and Brown, Michael E and Nesvorn{\`y}, David},
  journal={The Astrophysical Journal Letters},
  volume={966},
  number={1},
  pages={L8},
  year={2024},
  publisher={IOP Publishing}
}

@article{BMB24,
  title={A Pan-STARRS1 Search for Planet Nine},
  author={Brown, Michael E and Holman, Matthew J and Batygin, Konstantin},
  journal={The Astronomical Journal},
  volume={167},
  number={4},
  pages={146},
  year={2024},
  publisher={IOP Publishing}
}

@article{rice2020,
  title={Exploring Trans-Neptunian Space with TESS: A Targeted Shift-stacking Search for Planet Nine and Distant TNOs in the Galactic Plane},
  author={Rice, Malena and Laughlin, Gregory},
  journal={The Planetary Science Journal},
  volume={1},
  number={3},
  pages={81},
  year={2020},
  publisher={IOP Publishing}
}

@article{meisner2017,
  title={Searching for Planet Nine with coadded wise and neowise-reactivation images},
  author={Meisner, Aaron M and Bromley, Benjamin C and Nugent, Peter E and Schlegel, David J and Kenyon, Scott J and Schlafly, Edward F and Dawson, Kyle S},
  journal={The Astronomical Journal},
  volume={153},
  number={2},
  pages={65},
  year={2017},
  publisher={IOP Publishing}
}

@article{meisner2018,
  title={A 3$\pi$ Search for Planet Nine at 3.4 $\mu$m with WISE and NEOWISE},
  author={Meisner, AM and Bromley, BC and Kenyon, SJ and Anderson, TE},
  journal={The Astronomical Journal},
  volume={155},
  number={4},
  pages={166},
  year={2018},
  publisher={IOP Publishing}
}

@article{dlFM2022nodaldistances,
  title={Twisted extreme trans-Neptunian orbital parameter space: statistically significant asymmetries confirmed},
  author={De La Fuente Marcos, C and De La Fuente Marcos, R},
  journal={Monthly Notices of the Royal Astronomical Society: Letters},
  volume={512},
  number={1},
  pages={L6--L10},
  year={2022},
  publisher={Oxford University Press}
}

@article{bb23,
  title={On the proposed interstellar origin of the USG 20140108 fireball},
  author={Brown, Peter G and Borovi{\v{c}}ka, Ji{\v{r}}{\'\i}},
  journal={The Astrophysical Journal},
  volume={953},
  number={2},
  pages={167},
  year={2023},
  publisher={IOP Publishing}
}

@article{sn24,
  title={How likely is the interstellar origin of CNEOS14? On the reliability of the CNEOS database},
  author={Socas-Navarro, Hector},
  journal={arXiv preprint arXiv:2405.17219},
  year={2024}
}

@article{khain2018generation,
  title={The generation of the distant kuiper belt by planet nine from an initially broad perihelion distribution},
  author={Khain, Tali and Batygin, Konstantin and Brown, Michael E},
  journal={The Astronomical Journal},
  volume={155},
  number={6},
  pages={250},
  year={2018},
  publisher={IOP Publishing}
}

@article{brown2021orbit,
  title={The orbit of planet nine},
  author={Brown, Michael E and Batygin, Konstantin},
  journal={The Astronomical Journal},
  volume={162},
  number={5},
  pages={219},
  year={2021},
  publisher={IOP Publishing}
}

@article{siraj2024orbit,
  title={Orbit of a Possible Planet X},
  author={Siraj, Amir and Chyba, Christopher F and Tremaine, Scott},
  journal={arXiv preprint arXiv:2410.18170},
  year={2024}
}

@article{shankman2017ossos,
  title={OSSOS. VI. Striking biases in the detection of large semimajor axis trans-neptunian objects},
  author={Shankman, Cory and Kavelaars, JJ and Bannister, Michele T and Gladman, Brett J and Lawler, Samantha M and Chen, Ying-Tung and Jakubik, Marian and Kaib, Nathan and Alexandersen, Mike and Gwyn, Stephen DJ and others},
  journal={The Astronomical Journal},
  volume={154},
  number={2},
  pages={50},
  year={2017},
  publisher={IOP Publishing}
}

@article{gladman2021transneptunian,
  title={Transneptunian space},
  author={Gladman, Brett and Volk, Kathryn},
  journal={Annual Review of Astronomy and Astrophysics},
  volume={59},
  pages={203--246},
  year={2021},
  publisher={Annual Reviews}
}

@article{batygin2016evidence,
  title={Evidence for a distant giant planet in the solar system},
  author={Batygin, Konstantin and Brown, Michael E},
  journal={The Astronomical Journal},
  volume={151},
  number={2},
  pages={22},
  year={2016},
  publisher={IOP Publishing}
}

@article{gladman2002evidence,
  title={Evidence for an extended scattered disk},
  author={Gladman, Brett and Holman, M and Grav, T and Kavelaars, J and Nicholson, P and Aksnes, K and Petit, J-M},
  journal={Icarus},
  volume={157},
  number={2},
  pages={269--279},
  year={2002},
  publisher={Elsevier}
}

@article{tsiganis2005origin,
  title={Origin of the orbital architecture of the giant planets of the Solar System},
  author={Tsiganis, Kleomeris and Gomes, Rodney and Morbidelli, Alessandro and Levison, Hal F},
  journal={Nature},
  volume={435},
  number={7041},
  pages={459--461},
  year={2005},
  publisher={Nature Publishing Group UK London}
}

@article{levison2008origin,
  title={Origin of the structure of the Kuiper belt during a dynamical instability in the orbits of Uranus and Neptune},
  author={Levison, Harold F and Morbidelli, Alessandro and VanLaerhoven, Christa and Gomes, Rodney and Tsiganis, Kleomenis},
  journal={Icarus},
  volume={196},
  number={1},
  pages={258--273},
  year={2008},
  publisher={Elsevier}
}

@article{walsh2011low,
  title={A low mass for Mars from Jupiter’s early gas-driven migration},
  author={Walsh, Kevin J and Morbidelli, Alessandro and Raymond, Sean N and O'Brien, David P and Mandell, Avi M},
  journal={Nature},
  volume={475},
  number={7355},
  pages={206--209},
  year={2011},
  publisher={Nature Publishing Group UK London}
}

@article{raymond2014grand,
  title={The Grand Tack model: a critical review},
  author={Raymond, Sean N and Morbidelli, Alessandro},
  journal={Proceedings of the International Astronomical Union},
  volume={9},
  number={S310},
  pages={194--203},
  year={2014},
  publisher={Cambridge University Press}
}

@article{trujillo2014sedna,
  title={A Sedna-like body with a perihelion of 80 astronomical units},
  author={Trujillo, Chadwick A and Sheppard, Scott S},
  journal={Nature},
  volume={507},
  number={7493},
  pages={471--474},
  year={2014},
  publisher={Nature Publishing Group UK London}
}

@Article{H07,
  Author    = {Hunter, J. D.},
  Title     = {Matplotlib: A 2D graphics environment},
  Journal   = {Computing In Science \& Engineering},
  Volume    = {9},
  Number    = {3},
  Pages     = {90--95},
  publisher = {IEEE COMPUTER SOC},
  doi = {10.1109/MCSE.2007.55},
  year      = 2007
}

@article{numpy11, title={The NumPy array: a structure for
                  efficient numerical computation},
                  author={Van Der Walt, Stefan and Colbert, S Chris
                  and Varoquaux, Gael}, journal={Computing in Science
                  \& Engineering}, volume={13}, number={2},
                  pages={22--30}, year={2011}, publisher={AIP
                  Publishing}}

@Article{ipython07,
Author = {P\'erez, Fernando and Granger, Brian E.},
Title = {{IP}ython: a System for Interactive Scientific Computing},
Journal = {Computing in Science and Engineering},
Volume = {9},
Number = {3},
Pages = {21--29},
month = may,
year = 2007,
doi = {10.1109/MCSE.2007.53},

URL = { 
        https://aip.scitation.org/doi/abs/10.1109/MCSE.2007.53
    
},
eprint = { 
        https://aip.scitation.org/doi/pdf/10.1109/MCSE.2007.53}

}

@INPROCEEDINGS{NGD21,
       author = {{Napier}, K. and {Gerdes}, D. and {Dark Energy Survey Collaboration}},
        title = "{No Evidence for Orbital Clustering in the Extreme Trans-Neptunian Objects}",
    booktitle = {AAS/Division of Dynamical Astronomy Meeting},
         year = 2021,
       series = {AAS/Division of Dynamical Astronomy Meeting},
       volume = {53},
        month = jun,
          eid = {502.03},
        pages = {502.03},
       adsurl = {https://ui.adsabs.harvard.edu/abs/2021DDA....5250203N}
}

@article{brown2004discovery,
  title={Discovery of a candidate inner Oort cloud planetoid},
  author={Brown, Michael E and Trujillo, Chadwick and Rabinowitz, David},
  journal={The Astrophysical Journal},
  volume={617},
  number={1},
  pages={645},
  year={2004},
  publisher={IOP Publishing}
}

@article{brown2023MONDPlanet9,
  title={Modified Newtonian Dynamics as an Alternative to the Planet Nine Hypothesis},
  author={Brown, Katherine and Mathur, Harsh},
  journal={The Astronomical Journal},
  volume={166},
  number={4},
  pages={168},
  year={2023},
  publisher={IOP Publishing}
}

@ARTICLE{ScholtzUnwindBHPlanet9,
       author = {{Scholtz}, Jakub and {Unwin}, James},
        title = "{What if Planet 9 is a Primordial Black Hole?}",
      journal = {\prl},
         year = 2020,
        month = jul,
       volume = {125},
       number = {5},
          eid = {051103},
        pages = {051103},
          doi = {10.1103/PhysRevLett.125.051103},
archivePrefix = {arXiv},
       eprint = {1909.11090},
 primaryClass = {hep-ph},
       adsurl = {https://ui.adsabs.harvard.edu/abs/2020PhRvL.125e1103S}
}

@ARTICLE{BB21,
       author = {{Brown}, Michael E. and {Batygin}, Konstantin},
        title = "{The Orbit of Planet Nine}",
      journal = {\aj},
         year = 2021,
        month = nov,
       volume = {162},
       number = {5},
          eid = {219},
        pages = {219},
          doi = {10.3847/1538-3881/ac2056},
archivePrefix = {arXiv},
       eprint = {2108.09868},
 primaryClass = {astro-ph.EP},
       adsurl = {https://ui.adsabs.harvard.edu/abs/2021AJ....162..219B}
}

@ARTICLE{dLdlFMdlFM17,
       author = {{de Le{\'o}n}, J. and {de la Fuente Marcos}, C. and {de la Fuente Marcos}, R.},
        title = "{Visible spectra of (474640) 2004 VN$_{112}$-2013 RF$_{98}$ with OSIRIS at the 10.4 m GTC: evidence for binary dissociation near aphelion among the extreme trans-Neptunian objects}",
      journal = {\mnras},
         year = 2017,
        month = may,
       volume = {467},
       number = {1},
        pages = {L66-L70},
          doi = {10.1093/mnrasl/slx003},
archivePrefix = {arXiv},
       eprint = {1701.02534},
 primaryClass = {astro-ph.EP},
       adsurl = {https://ui.adsabs.harvard.edu/abs/2017MNRAS.467L..66D}
}

@ARTICLE{PASNS25,
       author = {{Pe{\~n}a-Asensio}, E. and {Socas-Navarro}, H. and {Seligman}, D.~Z.},
        title = "{Error dependencies in the space-based CNEOS fireball database}",
      journal = {\aap},
         year = 2025,
        month = sep,
       volume = {701},
          eid = {A202},
        pages = {A202},
          doi = {10.1051/0004-6361/202554224},
archivePrefix = {arXiv},
       eprint = {2508.01454},
 primaryClass = {astro-ph.EP},
       adsurl = {https://ui.adsabs.harvard.edu/abs/2025A&A...701A.202P}
}

\end{document}